\documentclass[fleqn,usenatbib]{mnras}

\usepackage{newtxtext,newtxmath}

\usepackage[T1]{fontenc}

\DeclareRobustCommand{\VAN}[3]{#2}
\let\VANthebibliography\thebibliography
\def\thebibliography{\DeclareRobustCommand{\VAN}[3]{##3}\VANthebibliography}

\usepackage{graphicx}	
\usepackage{amsmath}	
\usepackage{comment}

\title{A glitch in the millisecond pulsar J0900$-$3144}

\author[B. Bhat et al.]{Bhavnesh Bhat,$^{1}$\thanks{E-mail: bhavnesh.bhat@postgrad.manchester.ac.uk}
Michael J. Keith,$^{1}$\thanks{E-mail: michael.keith@manchester.ac.uk}
Isma\"el Cognard,$^{2,3}$
Lucas Guillemot,$^{2,3}$
Marcus E. Lower,$^{4}$
\newauthor
Matthew T. Miles,$^{5,6}$
Daniel J. Reardon,$^{4, 5}$
Golam Shaifullah,$^{7,8,9}$
Ryan M. Shannon,$^{4,5}$
Benjamin W. Stappers,$^{1}$
\newauthor
Gilles Theureau,$^{2, 3}$
Shuangqiang Wang,$^{10, 11}$
Andrew Zic,$^{4}$
Benjamin Shaw$^{1}$
\\
\small$^{1}$Jodrell Bank Centre for Astrophysics, Department of Physics and Astronomy, The University of Manchester, Manchester M13 9PL, UK\\
\small$^{2}$LPC2E, OSUC, Univ Orleans, CNRS, CNES, Observatoire de Paris, F-45071 Orleans, France\\
\small$^{3}$ORN, Observatoire de Paris, Universit\'e PSL, Univ Orl\'eans, CNRS, 18330 Nan\c{c}ay, France\\
\small$^{4}$Centre for Astrophysics and Supercomputing, Swinburne University of Technology, PO Box 218, Hawthorn, VIC 3122, Australia\\
\small$^{5}$OzGrav: The ARC Centre of Excellence for Gravitational Wave Discovery\\
\small$^{6}$Department of Physics and Astronomy, Vanderbilt University, 2301 Vanderbilt Place, Nashville, TN 37235, USA\\
\small$^{7}$Dipartimento di Fisica ``G. Occhialini'', Università degli Studi di Milano-Bicocca, \small Piazza della Scienza 3, 20126 Milano, Italy.\\
\small$^{8}$INFN, Sezione di Milano-Bicocca, Piazza della Scienza 3, 20126 Milano, Italy.\\
\small$^{9}$INAF - Osservatorio Astronomico di Cagliari, via della Scienza 5, 09047 Selargius (CA), Italy\\
\small$^{10}$CSIRO Astronomy and Space Science, PO Box 76, Epping, NSW 1710, Australia\\
\small$^{11}$Xinjiang Astronomical Observatory, Chinese Academy of Sciences, Urumqi, Xinjiang 830011, People's Republic of China \\
}

\date{Accepted XXX. Received YYY; in original form ZZZ}

\pubyear{\the\year{}}

\begin{document}
\label{firstpage}
\pagerange{\pageref{firstpage}--\pageref{lastpage}}
\maketitle

\begin{abstract}
We report the detection of a glitch in the millisecond pulsar (MSP) PSR J0900$-$3144, which is included in the European, MeerKAT and Parkes pulsar timing array experiments. The dataset combines observations from the MeerKAT, Nançay, Lovell, and Murriyang telescopes, spanning a total baseline of approximately 14 years. The glitch occurred on MJD~59942(17), with a measured fractional spin frequency step of $\Delta \nu_g / \nu=1.15(13) \times 10^{-12}$. This event represents the third glitch detected in a MSP, following those in PSRs~B1821$-$24A and J0613$-$0200. Although smaller in amplitude than the previous two, the glitch in PSR~J0900$-$3144 is of a comparable order of magnitude. The updated MSP glitch rate is $2.5(1)\times 10^{-3}$ glitches per pulsar per year, which suggests it is likely current PTAs will detect another MSP glitch within five years.
Using simulations, we demonstrate that such small glitches can go undetected, especially in short datasets such as those from new PTAs, and can bias the inferred achromatic noise model parameters, potentially leading to the down-weighting of the pulsar in gravitational wave background searches. 
\end{abstract}




\section{Introduction}
Pulsars are among the most stable astrophysical objects and are often referred to as ‘cosmic clocks’ due to the remarkable rotational stability that allows their pulse arrival times to be measured and modelled with remarkable precision. Millisecond pulsars (MSPs) are especially stable within the pulsar population and play a central role in pulsar timing array (PTA) experiments aimed at the statistical detection of the gravitational wave background (GWB) in the nanohertz frequency band~\citep{foster1990}. Several recent publications have reported evidence for the detection of the GWB~\citep{EPTA2023,NANOGrav2023,Miles2024,PPTA2023}.
The pulses from pulsars are recorded as times of arrival (ToAs), with deviations from a deterministic timing model referred to as \textit{timing residuals}. As the GWB signal introduces only subtle perturbations in the ToAs, it is essential to model all other sources of variation with high precision to avoid contaminating or masking the GWB signature. 
This includes accounting for both intrinsic and extrinsic noise sources which we collectively refer to as timing noise. 
Accurately modelling these contributions is crucial for improving the sensitivity of PTA experiments to gravitational waves.

The spin evolution of a pulsar is primarily governed by its spin-down rate, $\dot{\nu}$, which is assumed to be stable. However, pulsars occasionally exhibit sudden, irregular increases in spin frequency $\nu$, often accompanied by changes in $\dot{\nu}$. These abrupt events are known as \textit{glitches}. Glitches are believed to originate from internal dynamics within the neutron star, such as angular momentum transfer between the superfluid interior and the solid crust, and are commonly observed in young pulsars~\citep{Espinoza2011,Yu2013,Lower2021,basu2022}. In contrast, glitches in millisecond pulsars (MSPs) are rare. The first confirmed glitch in a MSP was reported by \citet{Cognard2004} in PSR B1821$-$24A, located near the core of the globular cluster M28. This event had a fractional frequency step of $\Delta \nu_g / \nu = 8(1) \times 10^{-12}$, making it one of the smallest glitches detected at the time. A second MSP glitch was subsequently discovered by \citet{Mckee2016} in PSR J0613$-$0200, based on data which was part of the European PTA Data Release 1. The glitch was characterized by a fractional frequency step of $\Delta \nu_g / \nu = 2.5(1) \times 10^{-12}$ and a fractional change in spin-down rate of $\Delta \dot{\nu_g} / \dot{\nu} = 1.6(39) \times 10^{-4}$.

The physical mechanism behind glitches remains an open question, but the prevailing theory attributes them to the transfer of angular momentum from the superfluid component of the neutron star's interior to its solid crust \citep{Haskell2018,Basu2018,Antonopoulo2022}. It is thought that the crust and the neutron superfluid are weakly coupled through mutual frictional forces. As the crust slows down due to the gradual loss rotational kinetic energy, the superfluid retains its rotational speed, leading to a lag between the two components. This lag induces a Magnus force~\citep{Sonin1997} that increases the stress on pinned superfluid vortices within the crust. Once this stress surpasses a critical threshold, the vortices unpin abruptly, allowing the superfluid to transfer angular momentum to the crust, thus spinning it up and resulting in a glitch~\citep{Chamel2013}. Another proposed mechanism for glitches involves starquakes, which occur as the neutron star’s crust adjusts from an oblate shape caused by rapid rotation toward a more spherical configuration as the star spins down~\citep{Baym1969,Espinoza2011}. However, recent studies argue that starquake based models are insufficient to account for the observed glitch sizes and glitch activity in the current broad radio pulsar population~\citep{Remcoret2021,Giliberti2019,Antonelli2022}.

A key challenge in detecting glitches, particularly those of small amplitude, arises from the presence of timing noise~\citep{Cordes1980}, which can obscure or mimic genuine changes in spin frequency and be absorbed in red noise over short data-span. This is especially problematic when attempting to distinguish a true glitch from variations caused by higher-order spin-down terms~\citep{shaw2018}. Timing noise manifests
as stochastic fluctuations in the pulse times of arrival (ToAs), often
attributed to intrinsic processes within the pulsar itself~\citep{Shannon2010}. In certain non-MSPs, quasi-periodic "noise" in the timing residuals have been observed, which are thought to arise from spin-down state switching ~\citep{Lyne2010,Brook2016,Shaw2022,Basu2024,Lower2025}, while in the MSPs it is typically modelled as a red noise process with a power-law spectrum~\citep{Coles2011,Lentati2013}.

PSR~J0900$-$3144 was discovered in the Parkes High-Latitude Pulsar Survey~\citep{Burgay2006}, with a spin period of 10\,ms and a spin-down rate of $\dot{P} = 4.9\times10^{-20}$\,s\,s$^{-1}$. It is in an 18.73-day binary orbit with a helium white dwarf (WD) companion of mass $\sim$0.38\,M$_\odot$~\citep{Chisabi2025}. Early observations using the MeerKAT telescope revealed anomalous timing behaviour suggestive of a possible glitch~\citep{Chisabi2025}. However, in the absence of sufficient statistical evidence, these variations were initially interpreted as red noise. In this work, we re-analyse the dataset in the context of extended multi-telescope observations, which now support a glitch interpretation. We also examine the broader implications for future glitch detections in PTA pulsars, emphasising the importance of robust noise characterisation when such events are present.

The structure of this paper is as follows: Section~\ref{section:Observation} describes the observational data and its processing. Section~\ref{section:timing} outlines the timing analysis and the modelling techniques employed. Section~\ref{sec:results} presents the results  of both the timing and Bayesian analyses. In Section~\ref{section:discussion}, we discuss the implications of our findings and the simulation done, and Section~\ref{sec:conclusion} summarises our conclusions.
\section{Observations}
\label{section:Observation}

\begin{figure*}
    \centering
    \includegraphics[width=\linewidth]{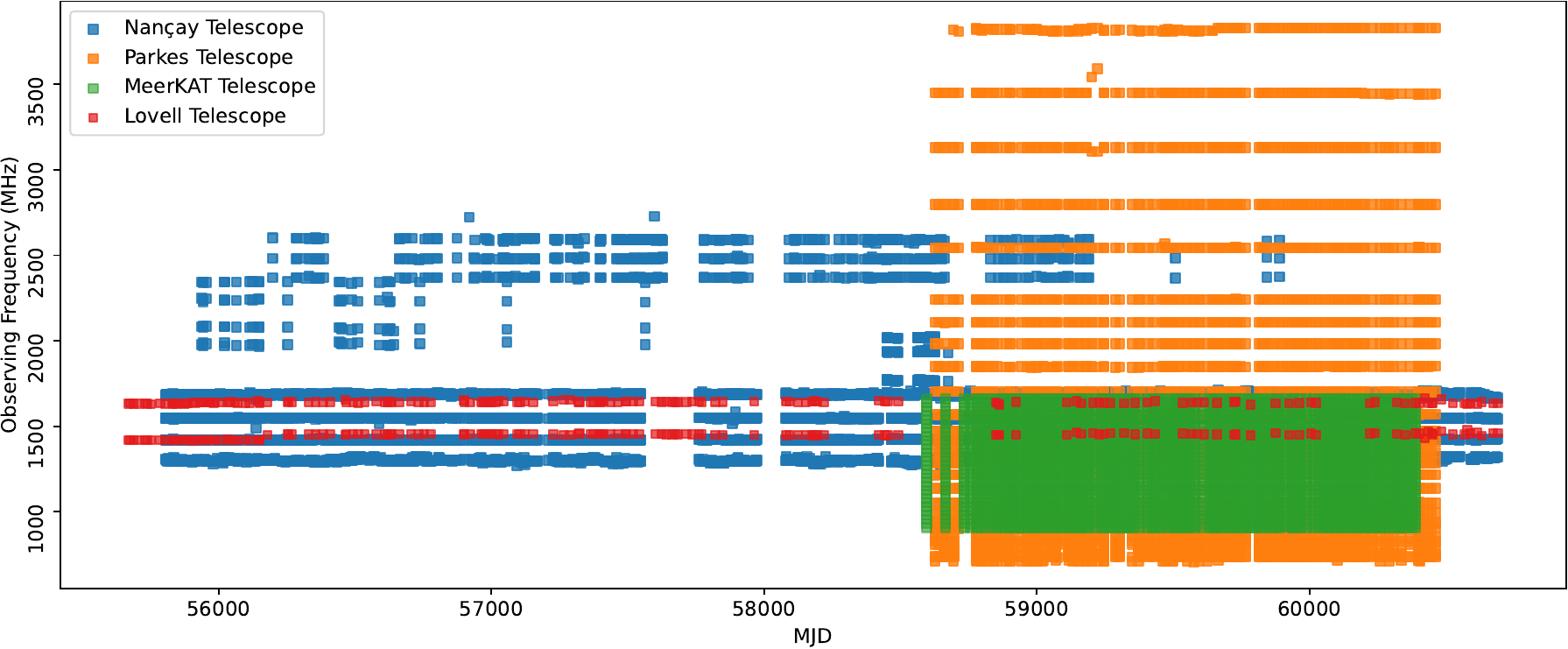}
    \caption{Frequency and time coverage of the different telescopes used in this study. Each point represents an individual time of arrival (ToA) measurement. Each observation has one ToA per sub-band, see Table~\ref{table:telescope_data} for further details.}
    \label{fig:fre_dist}
\end{figure*}

The dataset used in this work comprises observations from the MeerKAT Pulsar Timing Array (MPTA)~\citep{MPTA_data}, the Parkes Pulsar Timing Array (PPTA)~\citep{PPTA_data}, and the European Pulsar Timing Array (EPTA)~\citep{EPTA_2023_Data}, combining data from the MeerKAT, Murriyang, Nançay, and Lovell telescopes. These span a total time baseline of approximately 14 years (MJD 55669–60690). The frequency coverage and observing spans of each instrument are illustrated in Figure~\ref{fig:fre_dist}, with further details provided in Table~\ref{table:telescope_data}.

Observations with the Nançay Radio Telescope (NRT) were obtained using the Nançay Ultimate Pulsar Processing Instrument (NUPPI), which provides four central frequency observing bands, each further split into four sub-bands with a bandwidth of 512\,MHz per band. Integration times ranged from 20 to 80 minutes. The data were coherently dedispersed in real time. Template profiles were generated by summing the eight highest signal-to-noise (S/N) observations \citep{Guillemot2023}, and times of arrival (ToAs) were extracted using the matrix template matching (MTM) method implemented in \textsc{PSRCHIVE}~\citep{VanStraten2006}.

The Murriyang 64-m radio telescope observations were done using the Ultra-Wideband Low (UWL) receiver, with data acquisition via the Medusa signal processor \citep{Hobbs2020,PPTA_data}. The data were coherently dedispersed and folded in real time. Typical integration times were 3840\,s. ToAs were determined using the Fourier-domain Monte Carlo method via the \texttt{pat} tool in \textsc{PSRCHIVE}, using wideband portrait templates \citep{Pennucci2014}. Wide-band profile portraits are formed by creating 1D templates for each channel by evaluating the wide-band model at the channel centre frequency \citep{Curylo2023}. The UWL data is divided into sub-bands with the bandwidth of each sub-band increasing with the central frequency.

MeerKAT observations were conducted using the L-band receiver and recorded by the Pulsar Timing User Supplied Equipment (PTUSE) backend \citep{Bailes2020}. Data were coherently dedispersed in real time at the fiducial dispersion measure (DM) and folded at the topocentric spin period. Typical observation durations were 2048\,s. The data were processed using the MeerTime pipeline (\textsc{Meerpipe}), with ToAs also generated using the Fourier-domain Monte Carlo method and portrait-mode templates \citep{Pennucci2014}.

Observations from the Lovell Telescope at Jodrell Bank Observatory were recorded using the ROACH backend~\citep{Bassa2016}. Integration times varied from 10 to 55 minutes. Data were coherently dedispersed and folded in real time into 1024 phase bins. ToAs were extracted using the Fourier-domain Monte Carlo algorithm implemented in \texttt{pat} \citep{Verbiest2016}.

\begin{table*}
    \centering
    \caption{\label{table:telescope_data}Summary of the timing data for PSR~J0900$-$3144 used in this study. The table lists the contributing telescopes, their central observing frequencies ($f_{\mathrm{central}}$), total number of time-of-arrival (ToA) measurements, MJD ranges, number of frequency sub-bands ($n_{\mathrm{sub}}$), total bandwidth, and Median ToA error($\mu s$).}
    \begin{tabular}{||c|c|c|c|c|c|c||}
    \hline
        Telescope & $f_\mathrm{central}$ MHz  & $N_\mathrm{TOAs}$ & MJD Range & $n_\mathrm{sub}$ & Bandwidth (MHz) &Median ToA Error ($\mu s)$ \\
        \hline
        NRT & 1484 & 7373 & 55805$-$60690 & 4 & 512 & 2.044\\
               & 1854 & 180 & 58449$-$58674 & 4 & 512 & 1.714\\
               & 2154 & 97 & 55938$-$57564 & 4 & 512 & 2.337\\
               & 2539 & 634 & 56197$-$59889 & 4 & 512 & 4.015\\
        \hline
        MeerKAT & 1307 & 3679 & 58595$-$60387 & 32 & 776 & 3.18\\
        \hline
        Murriyang & 2368 & 3738 & 58626$-$60463 & 32 & 3328 & 4.54\\
        \hline
        Lovell & 1520 &  330 & 55669$-$60690 & 2 &  400 & 11.21\\
        \hline
    \end{tabular}
\end{table*}

\section{Timing Analysis}
\label{section:timing}
Timing analysis was initially performed using \textsc{tempo2} \citep{TEMPO2A}, starting from the timing ephemeris published as part of the MPTA 4.5-year data release \citep{Miles2024}
This ephemeris includes the pulsar spin parameters, position, proper motion, and binary parameters. The updated ephemeris after the analysis is given in Table \ref{tab:eph}.

Combining the data from the four instruments requires additional model parameters, to account for the differences in ToA generation and frequency coverage between the datasets.
Both the MeerKAT and Parkes ToAs make use of a pulse portrait 2-D template, however, there is an arbitrary phase difference between the templates \citep{Pennucci2014}, as well as a unknown difference in the frequency dependence when making the portrait that leads to an apparent DM offset between the two datasets.

\begin{figure*}
    \centering
    \includegraphics[width=0.99\linewidth]{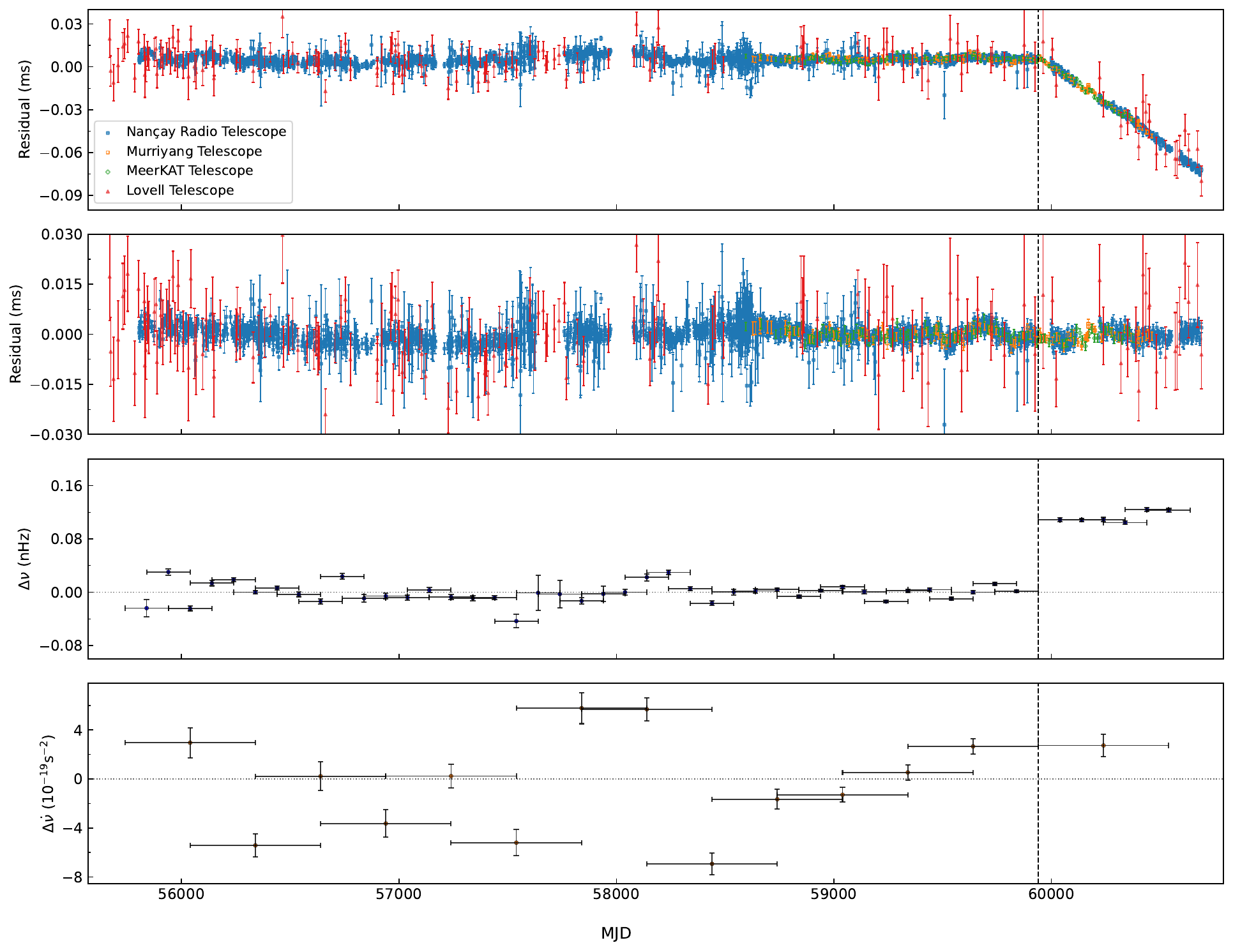}
    \caption{
    Comparison of timing residuals and spin parameter evolution for PSR~J0900$-$3144. 
    The top panel shows the frequency averaged pre-fit residuals, weighted by the uncertainty in ToA and the white noise parameters for each observing system for each observing epoch, clearly highlighting the glitch event. 
    The second panel presents the frequency averaged uncertainty-weighted post-fit residuals after subtracting the glitch model. 
    The third and fourth panels show an estimate of the deviations from the model pulsar spin frequency and frequency derivative, as obtained from fitting a quadratic to the short sections of the residuals in the top panel. $\Delta{\nu}$ is computed over 200-day windows, and $\Delta\dot{\nu}$ over 600-day windows.
    The vertical dashed line marks the glitch epoch.}
    \label{fig:residual_plot}
\end{figure*}

\begin{table}
\caption{Maximum likelihood timing parameters for PSR~J0900$-$3144. Errors in parentheses are $1$-$\sigma$ uncertainties estimated from the standard deviation of the marginalized posterior distributions (when fitted with \textsc{enterprise}), or from the least-squares fit in \textsc{tempo2}.}
\resizebox{\columnwidth}{!}{%
\begin{tabular}{ll}
\hline\hline
\multicolumn{2}{c}{Fit and data-set} \\
\hline
Pulsar name\dotfill & J0900$-$3144 \\ 
MJD range\dotfill & 55669.8$-$60690.1 \\ 
Data span,(yr)\dotfill & 13.74 \\ 
Number of TOAs\dotfill & 16031 \\
RMS timing residual,($\mu$s)\dotfill & 3.4 \\
\hline
\multicolumn{2}{c}{Spin Parameters} \\ 
\hline
Spin frequency,  (s$^{-1}$)\dotfill & 90.011841775793(2) \\ 
First derivative of spin  frequency, (s$^{-2}$)\dotfill & $-$3.959(2)$\times 10^{-16}$ \\ 
Glitch Epoch (GLEP, MJD) \dotfill & 59942.4(170) \\ 
Glitch spin frequency increment, $\Delta \nu_g$ (s$^{-1}$)  \dotfill & 1.04(12)$\times 10^{-10}$ \\
Glitch frequency derivative increment, $\Delta \dot{\nu}_g$ (s$^{-2}$)  \dotfill & 2.5(31)$\times 10^{-19}$ \\
\hline
\multicolumn{2}{c}{Astrometry and Binary Model} \\
\hline
Right ascension, (hh:mm:ss)\dotfill &  09:00:43.95218(1) \\ 
Declination,(dd:mm:ss)\dotfill & $-$31:44:30.8722(15) \\ 
Proper motion in right ascension, (mas\,yr$^{-1}$)\dotfill & $-$1.04(3) \\ 
Proper motion in declination, (mas\,yr$^{-1}$)\dotfill & 1.96(4) \\ 
Parallax, (mas)\dotfill & 1.4(1) \\ 
Orbital period,  (d)\dotfill & 18.7376360581(1) \\ 
Projected semi-major axis of orbit, (lt-s)\dotfill & 17.24880922(5) \\ 
Time of ascending node (MJD)\dotfill & 59105.639456390(9) \\ 
EPS1\dotfill & 9.891(6)$\times 10^{-6}$ \\ 
EPS2\dotfill & 3.495(6)$\times 10^{-6}$ \\ 
\hline
\multicolumn{2}{c}{Dispersion Model} \\
\hline
Dispersion measure, DM (cm$^{-3}$pc)\dotfill & 75.690(5) \\ 
First derivative of DM, (cm$^{-3}$pc\,yr$^{-1}$)\dotfill & $-$4(7)$\times 10^{-5}$ \\ 
$A_\mathrm{DM}$ \dotfill &$-$11.06(5) \\
$\gamma_\mathrm{DM}$ \dotfill &1.5 (3) \\
\hline
\multicolumn{2}{c}{Red Noise Gaussian Process} \\
\hline
$A_\mathrm{red}$  \dotfill &$-$12.62(4)\\
$\gamma_\mathrm{red}$\dotfill &1.6 (2) \\
\hline
\multicolumn{2}{c}{Chromatic Noise Gaussian Process} \\
\hline
$A_\mathrm{CHM}$ \dotfill &$-$13.4(1) \\
$\gamma_\mathrm{CHM}$  \dotfill &1.5(3) \\
$\beta_\mathrm{CHM}$ \dotfill &5.7(4) \\
\hline
\end{tabular}}
\label{tab:eph}
\end{table}
These are resolved by adding a \texttt{JUMP} parameter for the phase shift, and \texttt{FDJUMPDM} for the DM offset.
The \texttt{FDJUMPDM} parameter in \textsc{tempo2} accounts for frequency-dependent dispersive delays between observing systems or sub-bands, modelling delays not captured by the standard DM fit due to difference in portraits of different back-end resulting in difference for the DM fit.
The EPTA timing of Lovell and NRT makes use of an arbitrary phase shift between each sub-band, rather than frequency dependent portraits. Hence, we simply add the same \texttt{JUMP} parameters per sub-band as in the EPTA Data Release 2 \citep{EPTA_2023_Data}. The combined ToAs are shown in Figure \ref{fig:residual_plot}.

We model the glitch using the usual model defined by the epoch of the glitch (\texttt{GLEP}), step change in spin frequency $\Delta \nu$, and step change in spin frequency derivative $\Delta \dot{\nu}$, where the post-glitch phase~\citep{Demianski1983} is given by
\begin{equation}
    \phi(t) = \Delta \nu_g \ (t-\texttt{GLEP})s_\mathrm{d} + \frac{1}{2} \Delta \dot{\nu}_g \ (t-\texttt{GLEP})^2s_\mathrm{d}^2,
    \label{eqn:glitch}
\end{equation}
where $s_\mathrm{d} =86400\,\mathrm{s\,day}^{-1}$ converts from days to seconds. Note that we do not fit for an arbitrary phase offset at the glitch, as commonly used in linearised timing models in e.g. \textsc{tempo2} \citep{TEMPO2B}, because we use a non-linear fit that treats \texttt{GLEP} as a free parameter.

In order to accurately characterise the glitch, it is essential to account for stochastic variations in spin frequency, dispersion measure, and other chromatic processes that have been observed in this pulsar \citep{EPTA_noise_ppr,MPTA_data}. 
We adopt the widely used Fourier-domain Gaussian process (GP) framework for PTA analysis, as described in \citet{Lentati2013}. 
The achromatic spin red noise is modelled using a power-law GP, and posterior parameter estimation is carried out using nested sampling as implemented in \texttt{dynesty} \citep{DynestySampelr}. 
The red noise power spectral density is described by a power-law:
\begin{equation}
P(f) = \frac{A_{\mathrm{red}}^2}{12 \pi^2} \left(\frac{f}{\mathrm{yr}^{-1}}\right)^{-\gamma_{\mathrm{red}}} \mathrm{yr}^3,
\label{eqn:red_noise}
\end{equation}
where $A_{\mathrm{red}}$ is the red noise amplitude and $\gamma_{\mathrm{red}}$ is the spectral exponent.

In addition to achromatic red noise, which is believed to be intrinsic to the pulsar rotation \citep{Shannon2010}, there are also `chromatic', i.e., observing frequency dependent, processes due to propagation through the ionised interstellar medium. Primarily this arises from variations in the DM, which scales with observing frequency as $f_\mathrm{obs}^{-2}$.
Similarly to the achromatic noise, we model the DM varaitions as a Gaussian process characterised by a power-law amplitude $A_{\mathrm{DM}}$ and exponent $\gamma_{\mathrm{DM}}$ \citep{Lentati2013B}.
We also consider a more general chromatic noise process, with an arbitrary frequency dependence given by $f_\mathrm{obs}^{-\beta_{\mathrm{CHM}}}$ \citep{MPTA_data}, to account for chromatic process that are not captured by the DM variations alone \citep{Boris2021}.
This chromatic noise is likewise described using a power-law Gaussian process, with amplitude $A_{\mathrm{CHM}}$ and spectral exponent $\gamma_{\mathrm{CHM}}$, and defined such that the Gaussian process describes the signal at a reference frequency of 1400~MHz.

For white noise modelling, we include  the parameters \texttt{EFAC} accounting for miscalibration of the radiometer noise, \texttt{EQUAD} to represent additional time independent white noise, and an additional \texttt{ECORR} term, which accounts for correlated noise between ToAs observed within the same epoch \citep{vanHaasteren2014,TPA_whitenoise}:

\begin{equation}
\mathcal{\textbf{C}} = \mathtt{EFAC}^2 \ \Sigma + \mathtt{EQUAD}^2 \ \textbf{I} + \mathtt{ECORR}^2 \ \Delta
\end{equation}
where, $\Sigma_i=\sigma_i^2$ is a diagonal matrix, \textbf{I} is the identity matrix and $\Delta_{i,j}=1$ is the block diagonal matrix, for i,j is 1 for same epoch and 0 for the other case.

We use \textsc{enterprise} \citep{EnterpriseOrg}, via the \textsc{run\_enterprise} interface \citep{run_enterprise}, to infer the noise hyper-parameters ($A_\mathrm{red}, A_\mathrm{DM}, A_\mathrm{CHM}$, $\gamma_\mathrm{red}, \gamma_\mathrm{DM}, \gamma_\mathrm{CHM}$, $\beta_\mathrm{CHM}$), which describe the statistical properties of the stochastic noise processes quantified by Equation~\ref{eqn:red_noise}, as well as the glitch parameters (\texttt{GLEP}, $\Delta\nu$, $\Delta\dot\nu$), and white noise parameters (\texttt{EFAC}, \texttt{EQUAD} and \texttt{ECORR} for each instrument), while marginalising over all remaining deterministic timing model parameters.

\section{Results}
\label{sec:results}

The results of the analysis are presented in Table~\ref{tab:eph}, and posterior distributions are provided in Appendix~\ref{appendix}. The properties of the chromatic noise are consistent with those observed in the MPTA 4.5-year data set \cite[][]{MPTA_data}. In contrast to that analysis, we identify significant stochastic dispersion measure variations, which we attribute to the longer data span of our observations.

We also repeated the Bayesian analysis with a `no-glitch' model, that excludes the glitch but contains all other terms, and find a log Bayes factor $\Delta\ln Z \sim 15$ in favour of the model that included the glitch. 
This can be interpreted as strong evidence in support of the glitch~\citep{Kass1995,Parthasarathy2019,Lower2021}, and hence we only consider the results for the model which includes the glitch.

The glitch parameters derived from the complete model (i.e. including the glitch) correspond to a fractional frequency step of $\Delta\nu_g / \nu = 1.15(13) \times 10^{-12}$ and a fractional spin-down rate change of $\Delta\dot{\nu}_g / \dot{\nu} = -6.3(79) \times 10^{-4}$. We note that the best-fit $\Delta\dot{\nu}_g$ is positive, unlike in most observed glitches, however given the large uncertainty we consider it consistent with zero.
These values are comparable in scale to previous MSP glitches~\citep{Mckee2016,Cognard2004}, suggesting a typical range for such events.

In addition to the post-fit timing residuals, Figure~\ref{fig:residual_plot} shows estimates of the temporal evolution of $\nu$ and $\dot\nu$ relative to the pre-glitch timing model. 
These quantities were obtained by fitting subsets of the frequency-averaged pre-fit residuals (first panel of Figure~\ref{fig:residual_plot}) with a quadratic function, using sliding 200-day windows for $\Delta\nu$, and 600-day windows for $\Delta\dot{\nu}$. 
Whilst we do subtract the chromatic processes from the residuals, the achromatic spin noise is not removed, and the errors in these $\Delta\nu$ and $\Delta\dot{\nu}$ estimates do not include any contribution from the red noise.
The $\Delta\nu$ time-series clearly shows the distinct step in spin frequency caused by the glitch.
In contrast, the $\Delta\dot{\nu}$ time series seems largely unchanged by the glitch, showing only a scatter of order $4\times10^{-19}\,\textrm{s}^{-2}$ which we attribute to the achromatic red noise.
When accounting for the timing noise, the estimates of $\Delta\dot{\nu}$ are consistent with zero.
 
\subsection{MSP Glitch Rate}
\label{sec:glitch_rate}
With three confirmed glitches now known in MSPs, we can update previous rate estimates from \citet{Cognard2004} and \citet{Mckee2016}. Since those studies, the number of MSPs monitored, particularly in PTA experiments has increased substantially. Using the latest PTA data releases~\citep{EPTA_2023_Data,Nanograv_data,PPTA_data,MPTA_data}, we identify a total of 122 unique MSPs observed over spans exceeding four years, giving a cumulative observing time of approximately 1192 pulsar-years.

From this, we estimate an updated MSP glitch rate of $2.5 \pm 1.4$ glitches per 1000 pulsar-years, corresponding to one glitch per $\sim400$ years for a single MSP. Assuming this is the intrinsic glitch rate which is constant for all the MSP, we use Poisson statistics to forecast the probability of future detections of glitches. For the current IPTA (including MPTA) monitoring programme, the probability of observing at least one glitch in the next 5 years is approximately 78 percent, rising to 99 percent over a 15-year period.

\section{Discussion}
\label{section:discussion}
Given the increasing likelihood of glitches appearing in future PTA datasets, we find it worthwhile to consider the effect of unmodelled glitches on the PTA noise modelling.
This is especially important given the number of short datasets for the new pulsars being added to the PTA programmes, where distinguishing glitches from timing noise can be difficult~\citep{shaw2018}.

\begin{figure}
    \centering
    \includegraphics[width=\linewidth]{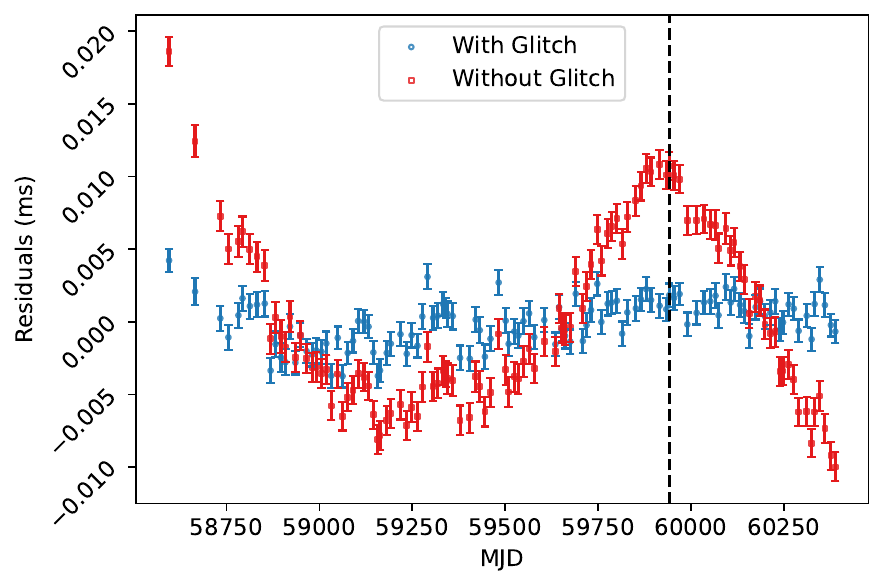}
    \caption{Post-fit timing residuals from the MeerKAT-only dataset, with and without including the glitch in the model. For clarity, the residuals from each observation have been combined to a single frequency-averaged residual, weighted by the ToA uncertainty and the white noise parameters.}
    \label{fig:meerkat_residuals}
\end{figure}

Indeed, despite the high sensitivity and broad bandwidth, if we use only the 5-year MeerKAT dataset, the Bayesian analysis is barely able to distinguish the glitch from achromatic red noise, with $\ln Z \sim 2.6$ in favour of the glitch model. As seen in Figure~\ref{fig:meerkat_residuals}, the model without the glitch is hard to distinguish from an achromatic red noise process. This is primarily due to the short observational baseline, which may allow the glitch signature to be partially absorbed into the timing model, particularly through higher-order spin-down terms.

\begin{figure}
    \centering
    \includegraphics[width=\linewidth]{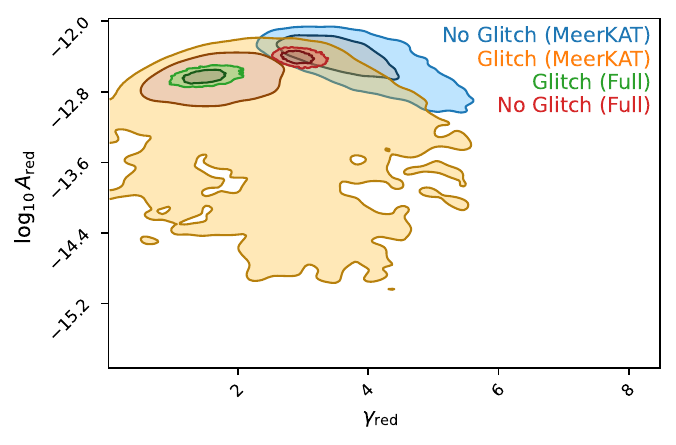}
    \caption{Comparison of red noise parameter posteriors between glitch and non-glitch models for different datasets. The unmodelled glitch has the effect of biasing both amplitude and spectral exponent of red noise.}
    \label{fig:rn_comp_alp}
\end{figure}

The impact of glitch modelling on red noise characterisation is illustrated in Figure~\ref{fig:rn_comp_alp}, which presents the posterior distributions of the red noise parameters $A_{\mathrm{red}}$ and $\gamma_{\mathrm{red}}$. A clear discrepancy is visible between the glitch and non-glitch models, for both the 5-year MeerKAT-only dataset and the longer combined dataset. In the absence of glitch modelling, the red noise parameters are biased, with the glitch signal being partially absorbed into the stochastic noise component. This results in a mischaracterisation of the red noise, most notably in the amplitude ($A_{\mathrm{red}}$) and spectral exponent ($\gamma_{\mathrm{red}}$), demonstrating the importance of accounting for glitches when modelling pulsar noise processes.

However, the issues raised by the model mis-specification is only problematic if we are unable to detect the glitch.
To investigate this, we have produced simulated datasets containing glitches at approximately 0.1, 0.5, 1, 2 and 4 times the size of the glitch in PSR~J0900$-$3144.
The simulated ToAs are broadly based on the MPTA observations of PSR~J0900$-$3144, however to make the analysis computationally feasible we simulate only frequency-averaged ToAs, and exclude any chromatic processes from the analysis.
The ToAs are simulated using \texttt{tempo2} with a cadence of 10 days, and white noise rms of 10$^{-3}$ $\mu$s.
The white noise level was chosen to be typical of the error in the observed frequency averaged ToA after accounting for the chromatic processes.
Red noise was injected using the \texttt{addRedNoise} plug-in, with hyper-parameters set to the post-fit values listed in Table~\ref{tab:eph}.

We simulated 25-year datasets for each of four glitch sizes with the glitch epoch at the mid-point.
We then use \textsc{run\_enterprise} to analyse the simulated data in a similar way to the real data.
We compare two models, the `glitch model', which fits for a glitch, plus red and white noise, and the `no glitch model' which only fits for the red and white noise.
In both cases we also marginalise over the pulsar timing model.
We perform the analysis on the full dataset, and shorter sub-sets, recording both the red noise model hyper-parameters and the Bayesian evidence in favour the glitch model.
We repeat this process 13 times and take the mean and standard deviation of the results in order to capture  realisation-to-realisation variations.

\begin{figure}
    \centering
    \includegraphics[width=\linewidth]{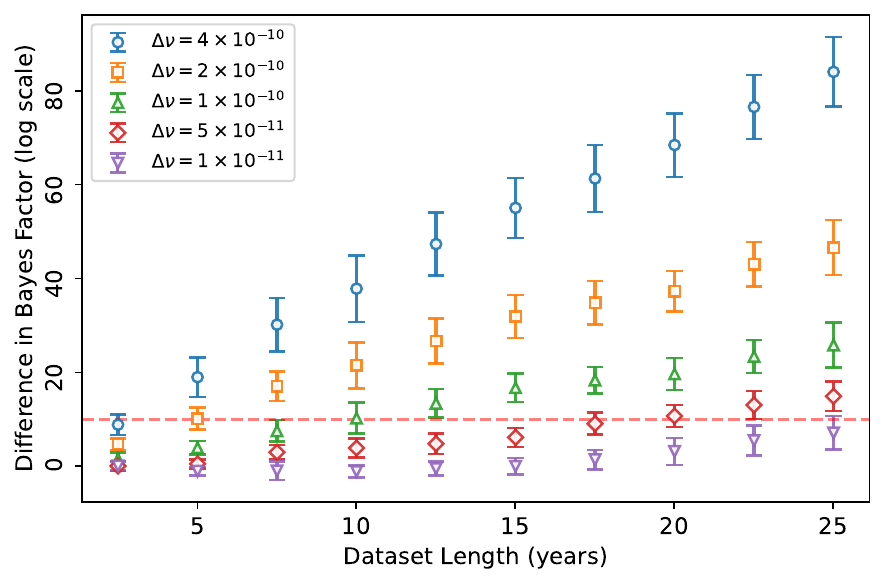}
    \caption{Bayesian evidence in favour of the glitch model as a function of total observation span and glitch amplitude. The horizontal dashed line indicates a Bayes factor of 10, broadly considered to be a significant detection.
    }
    \label{fig:glitch_work_simulation}
\end{figure}

\begin{figure}
    \centering
    \includegraphics[width=\linewidth]{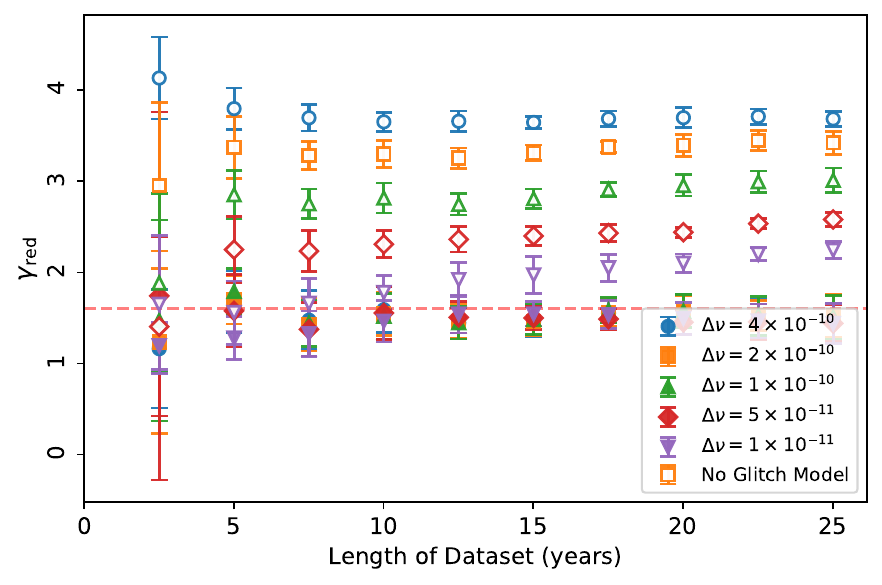}
    \caption{Comparison of the estimated $\gamma_{\mathrm{red}}$ between glitch and non-glitch models using \textsc{run\_enterprise}, for varying observation spans. Filled symbols are for the glitch model and open symbols are for the no-glitch model. The horizontal dashed line represent the injected $\gamma_\mathrm{red}$ modelled from the analysis of the PSR~J0900$-$3144 as in Table~\ref{tab:eph}. The systematic bias introduced by neglecting the glitch is clearly evident, with the non-glitch models consistently mis-estimating the red noise spectral exponent.} 
    \label{fig:gamma_red_all_n}
\end{figure}

Figure~\ref{fig:glitch_work_simulation} shows how the evidence for the glitch model varies with dataset length and glitch size.
This demonstrates that as expected, the ability to distinguish between a glitch and red noise increases with time.
Achieving a statistically significant detection (defined as $\ln Z \geq 10$) from Bayesian analysis typically requires an significant observational baseline depending on the size of the glitch as evident from the simulation shown in Figure~\ref{fig:glitch_work_simulation}. 
Small glitches in short datasets remain effectively indistinguishable from red noise, highlighting a key challenge in pulsar timing analyses.
Figure~\ref{fig:gamma_red_all_n} further illustrates the impact of failing to model a glitch on the inferred red noise spectral exponent, $\gamma_{\mathrm{red}}$. Across a range of injected glitch amplitudes, the no glitch model consistently overestimates $\gamma_{\mathrm{red}}$, as the glitch signal is partially absorbed into the red noise component. This leads to a steeper apparent power-law, misrepresenting the true noise structure. While the discrepancy is relatively minor for small glitches, it still induces deviation from the true parametric values potentially biasing the spectral exponent estimation. 

This has important implications for PTA analysis. While \citet{Mckee2016} examined a glitching PTA pulsar, they noted that only post glitch data has been used for PTA analysis. In contrast, our study includes a glitch that occurs mid-way through the MeerKAT-only dataset and near the end of the full data span, posing a greater challenge for accurate noise modelling. If the glitch is not properly modelled, its signature can be absorbed into the red noise component, leading to inflated estimates of red noise amplitude and spectral exponent. This, in turn, causes the pulsar to be down-weighted in PTA analyses, diminishing its contribution to common signal searches. Moreover, model mis-specification undermines the assumption of stationary, power-law Gaussian noise, potentially biasing parameter estimates or uncertainties in ways that are difficult to predict. These effects highlight the need for careful glitch modelling in PTA datasets to ensure robust and accurate inference.

\subsection{Comparison to the canonical pulsars}
It is beyond the scope of this paper to fully explore a comparison between MSP glitches and those of the canonical pulsars.
However, we can make some coarse comparison of the observed magnitude of the glitches in relation to $\nu$ and $\dot\nu$. 
\citet{basu2022} characterise the glitch activity of the canonical pulsars by $\dot{\nu}_g$, the product of the glitch rate and the average glitch size for a given pulsar.
They find that across the population, the mean glitch activity is best characterised by $\dot{\nu}_g / |\dot{\nu}| \simeq 0.018$.
Taking the glitch rate for MSPs derived in Section~\ref{sec:glitch_rate} of $2.5\times10^{-3}\,\textrm{yr}^{-1}$, we can use this relation to estimate an average glitch amplitude of around $10^{-7}\,\mathrm{Hz}$ for PSR~J0900$-$3144.
This is somewhat larger than the observed MSP glitch size, however, the relation for the mean activity in \citet{basu2022} seems to be dominated by the largest glitches, and the majority of glitches fall below the line, often by several orders of magnitude. Alternatively, the $\Delta\nu$–$\dot{\nu}$ relation reported by \citet{Espinoza2011} suggests an expected frequency step of $\Delta\nu \sim 10^{-10}\,\mathrm{s}^{-1}$, which is consistent with our measured glitch size.

A similar estimate can be made for $\Delta \dot\nu_g$, either by the relationship observed between $\Delta \dot\nu_g$ and $\Delta\nu_g$ for glitches (e.g. \citealp{Espinoza2011}), or by using the model for the glitch recovery component of $\ddot\nu$ as a function of $\nu$ and $\dot\nu$ in \citet{Yang2024}.
For the former, \citet{Espinoza2011} suggests that a glitch with $\Delta \nu_g \sim10^{-10}\,\mathrm{s}^{-1}$ would have $\Delta \dot\nu_g \sim -10^{-18} \mathrm{s}^{-2}$. However, we note that the apparent correlation shown in their diagram may be influenced by detection biases and cadence of ToAs, and it is not clear that the two quantities are intrinsically related~\citep{Watts2015,Antonopoulo2022}.
Alternatively if we assume that the glitch recovery $\ddot\nu$ is counteracted by the $\Delta \dot\nu_g$ of glitches, as believed to be the case for canonical glitching pulsars, we can use our estimated glitch rate to make a very coarse prediction for the typical size of $\Delta \dot\nu_g$.
For a glitch every 400 years with the $\nu$ and $\dot\nu$ of PSR~J0900$-$3144, this would also suggest $\Delta \dot\nu_g \sim -10^{-18}\,\mathrm{s}^{-2}$.
If we assume that the glitch in PSR~J0900$-$3144 really does have a negative $\Delta \dot\nu_g$, then our analysis suggests it is unlikely to be as large as $-10^{-18}\,\mathrm{s}^{-2}$, however both estimates of the expected size of $\Delta \dot\nu_g$ are based on relations with a lot of scatter, and a glitch of order $-10^{-19}\,\mathrm{s}^{-2}$ would be broadly compatible with both the data and the broad predictions of glitch sizes.
Further observations may help constrain this somewhat, but the error in the pre-glitch $\dot\nu$ is $2\times 10^{-19}\,\mathrm{s}^{-2}$, and is dominated by the achromatic red noise.
If we ignore the low frequency achromatic red noise in the estimation of $\dot\nu$, the theoretical uncertainty improves by two orders of magnitude, so things may improve somewhat as the noise process becomes better constrained. Nevertheless it may be hard to improve the constraint on $\Delta \dot\nu_g$ by more than an order of magnitude or so, even with more data.

The glitch in PSR~J0900$-$3144 is the smallest measured in any pulsar in both frequency step and fractional glitch size, and even appears small when the parameters are extrapolated from the relation derived for the canonical pulsars. However, given the broad range of glitch magnitudes we observe in the canonical pulsars, there does not seem to be any need for a separate population of glitches in the MSPs based on the glitch size alone.  

\section{Conclusions}
\label{sec:conclusion}
We report the detection of a glitch in PSR~J0900$-$3144, representing the smallest glitch observed in a MSP to date, yet comparable in magnitude to the two previously known MSP glitches. This detection has been statistically validated through Bayesian analysis of a combined dataset spanning nearly 14 years, incorporating observations from multiple telescopes. 

Our results demonstrate that glitches of such small amplitude become increasingly detectable as the observing baseline increases, with Bayesian model comparison favouring the glitch model significantly over longer timespans. In the absence of glitch modelling, the signal is absorbed into the achromatic red-noise component of the timing model, leading to systematic biases in the inferred red-noise parameters, particularly an overestimation of the spectral exponent $\gamma_{\mathrm{red}}$. This effect alters the shape of the inferred red-noise power spectrum, making it appear artificially steep.

To explore this further, we conducted targeted simulations across a range of glitch amplitudes and dataset lengths. These simulations confirm that unmodelled glitch signals are consistently absorbed into the red-noise model, especially for shorter baselines or smaller glitch steps. This has important implications for PTA experiments, as such mischaracterisation can reduce the reliability of noise modelling and down-weight the contribution of affected pulsars in GWB searches.

Finally, we updated the MSP glitch rate by incorporating this new detection into the growing PTA dataset. Our revised estimates suggest that additional MSP glitches are likely to be detected in the near future as datasets continue to grow in timespan and precision. These results reinforce the necessity of incorporating robust glitch modelling into PTA analyses to ensure accurate noise characterisation.
\section*{Acknowledgements}
The authors acknowledge  Avishek Basu for exciting discussion on glitches in Non-MSPs.
MEL is supported by an Australian Research Council (ARC) Discovery Early Career Research Award DE250100508.
Parts of this work were supported through the Australian Research Council Centre of Excellence for Gravitational Wave Discovery (CE170100007, CE230100016).
The authors acknowledge funding from the United Kingdom’s Research and Innovation (UKRI) Science and Technology Facilities Council (STFC), project reference [ST/Y509814/1]. 
Pulsar observation with the Lovell telescope are supported by funding from STFC.
The MeerKAT telescope is operated by the South African Radio Astronomy Observatory, which is a facility of the National Research Foundation, an agency of the Department of Science and Innovation.
Murriyang, CSIRO’s Parkes radio telescope, is part of the Australia Telescope National Facility (\href{https://ror.org/05qajvd42 [ror.org]}{https://ror.org/05qajvd42 [ror.org]}) which is funded by the Australian Government for operation as a National Facility managed by CSIRO. We acknowledge the Wiradjuri people as the Traditional Owners of the Observatory site.
The Nan\c{c}ay Radio Observatory is operated by the Paris Observatory, associated with the French Centre National de la Recherche Scientifique (CNRS). We acknowledge financial support from the ``Action Th\'ematique de Cosmologie et Galaxies'' (ATCG), ``Action Th\'ematique Gravitation R\'ef\'erences Astronomie M\'etrologie'' (ATGRAM) and ``Action Th\'ematique Ph\'enomènes Extr\^emes et Multi-messagers'' (ATPEM) of CNRS/INSU, France. 

\section*{Data Availability}
The observations of PSR J0900$-$3144 used in this work are part of the next data releases from the MPTA, PPTA and EPTA collaborations, and a combined dataset will appear in the Third Data Release from the International Pulsar Timing Array, which is expected to be published within 12 months of the submission of this article.
Authors may request access to the data used in this work by contacting the corresponding author, but are strongly recommended to make use of these official data releases once available.



\bibliographystyle{mnras}
\bibliography{reference} 



\appendix

\section{Posterior Distributions for Glitch and Noise Parameters}
\label{appendix}

In this appendix, we present the posterior distributions of the parameters inferred using \textsc{run\_enterprise} applied to the full dataset, incorporating both noise and glitch modelling. The noise model includes contributions from achromatic red noise, chromatic noise, dispersion measure variations, and white noise components. The glitch model accounts for a permanent step in spin frequency ($\Delta \nu$) and spin-down rate ($\Delta \dot{\nu}$), as well as the glitch epoch.

\begin{figure*}
    \centering
    \includegraphics[width=\linewidth]{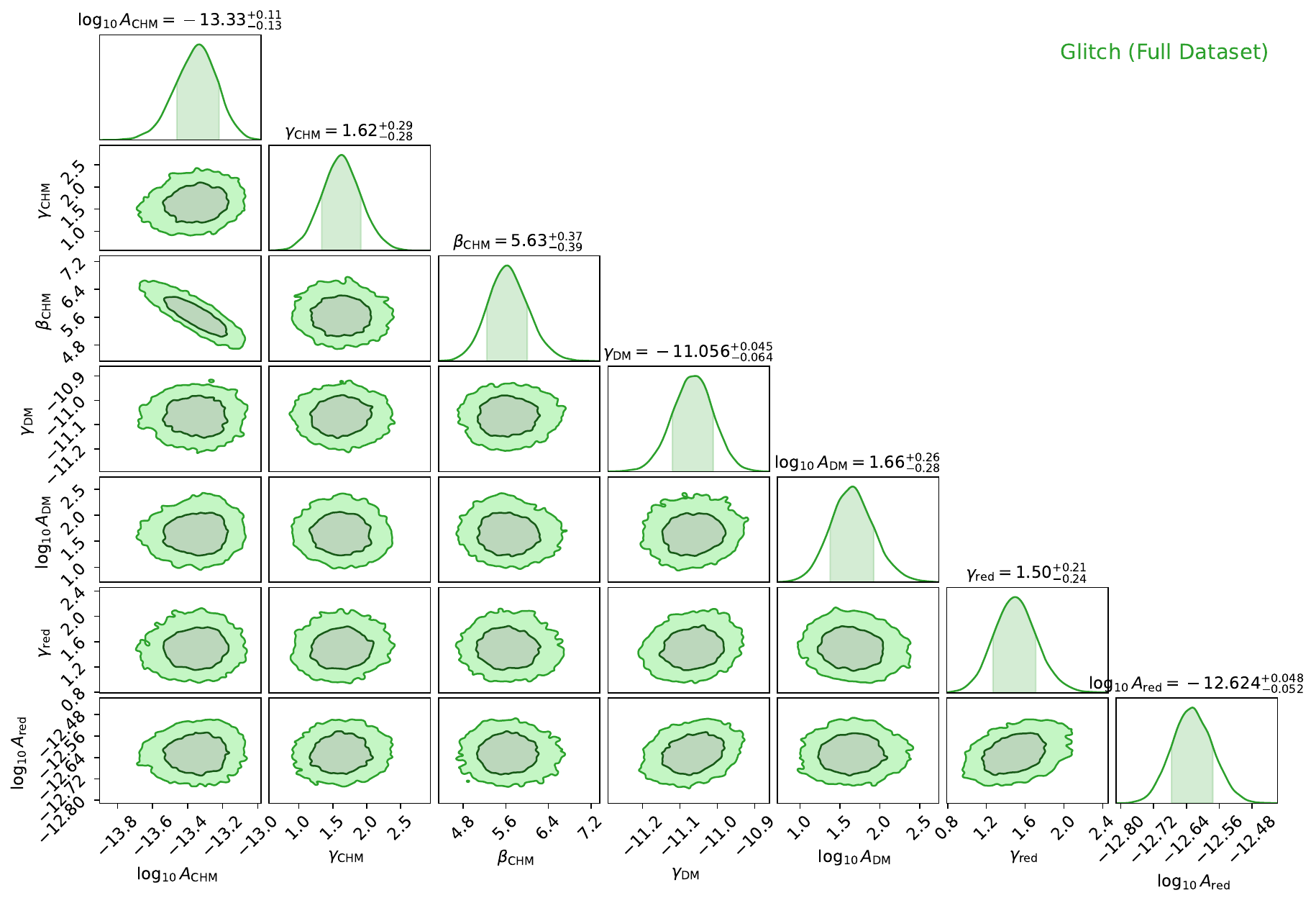}
    \caption{Posterior distributions of the noise model parameters obtained from the full dataset analysis. These include red noise, chromatic noise, and DM noise components used in the full Bayesian timing model.The shaded region indicates the 68\% and 95\% confidence intervals of the posterior distribution.}
    \label{fig:noise_corner_plot}
\end{figure*}

Figure~\ref{fig:noise_corner_plot} shows the posterior distributions for the noise parameters, including the chromatic noise amplitude and spectral exponent ($A_\mathrm{CHM}$, $\gamma_\mathrm{CHM}$), chromatic index of the noise process ($\beta_\mathrm{CHM}$), DM noise amplitude and spectral exponent ($A_\mathrm{DM}$, $\gamma_\mathrm{DM}$), and red noise amplitude and spectral exponent ($A_\mathrm{red}$, $\gamma_\mathrm{red}$). These components represent the long-term stochastic processes influencing the timing residuals.

\begin{figure*}
    \centering
    \includegraphics[width=\linewidth]{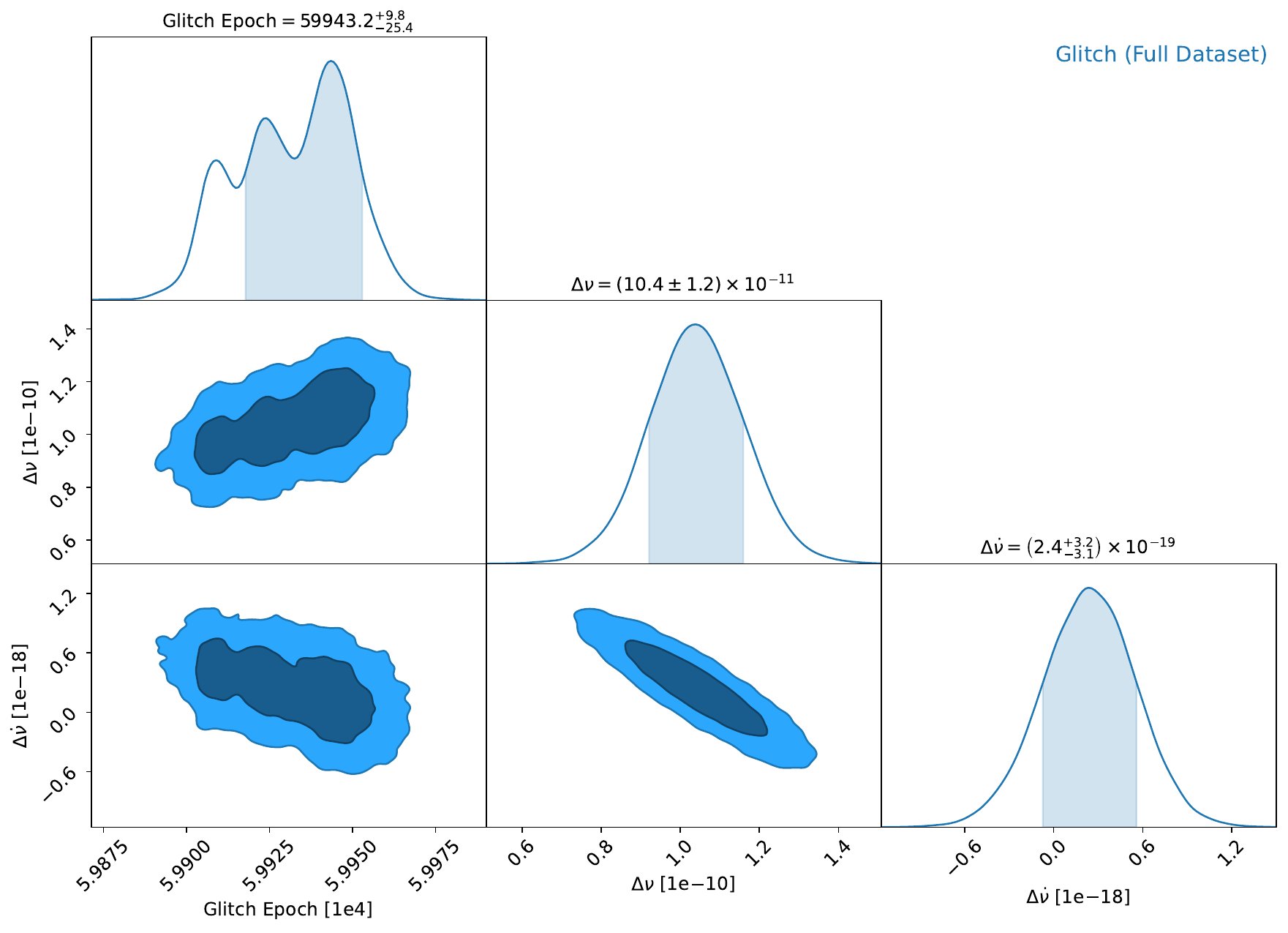}
    \caption{Posterior distributions of the glitch parameters derived from the full dataset analysis. Shown are the inferred glitch epoch (MJD), spin frequency step $\Delta \nu$, and spin-down rate step $\Delta \dot{\nu}$. The shaded region indicates the 68\% and 95\% confidence intervals of the posterior distribution.}
    \label{fig:glitch_corner_plot}
\end{figure*}

Figure~\ref{fig:glitch_corner_plot} shows the posterior distributions for the glitch parameters: glitch epoch (MJD), spin frequency step ($\Delta \nu$), and spin-down rate step ($\Delta \dot{\nu}$). The tight constraints on these parameters indicate a statistically significant glitch detection in PSR~J0900$-$3144.

\bsp	
\label{lastpage}
\end{document}